\documentstyle[aps,prb,twocolumn,floats,psfig]{revtex}

\newcommand{\Y}{YNi$_2$B$_2$C~}
\newcommand{\Lu}{LuNi$_2$B$_2$C~}

\newcommand{\vk}{\bf k\rm}

\newcommand{\vH}{\bf H\rm}

\newcommand{\De}{$\Delta(\vk)~$}

\begin{document}

\draft

\title{Thermal conductivity in superconducting borocarbides \Lu and \Y
\cite{Conf02}}

\author{P. Thalmeier$^1$ and K. Maki$^{2,3}$}

\address{$^1$Max-Planck-Institute for the Chemical Physics of Solids,
N\"othnitzer Str.40, 01187 Dresden, Germany}
\address{$^2$Max-Planck-Institute for the Physics of Complex Systems,
N\"othnitzer Str.38, 01187 Dresden, Germany}
\address{$^3$Department of Physics and Astronomy,
University of Southern California, Los Angeles,}
\address{CA 90089-0484, USA}

\maketitle

\begin{abstract}
We have recently proposed the s+g wave model for superconducting
borocarbides. In spite of a substantial s-wave component, this order
parameter exhibits the $\sqrt{H}$ dependent specific heat and a
thermal conductivity linear in H in the vortex state. This is
characteristic for nodal superconductors when T,$\Gamma\ll\Delta$
where $\Gamma$ is the quasiparticle scattering rate and $\Delta$ the
maximum superconducting gap. Here we investigate the thermal
conductivity parallel to the c- and a- axis in a magnetic field
tilted by $\theta$ from the c- axis and rotating within the a-b plane.
\end{abstract}


The superconductivity in the rare earth borocarbides \Lu
and \Y is of great interest\cite{Canfield98,Mueller01}. We have
proposed recently the superconducting order parameter
\cite{Maki02,Jang02} 

\begin{equation}
\label{GAP}
\Delta(\vk)=\frac{1}{2}\Delta(1+\sin^4\vartheta\cos(4\varphi))
\end{equation}

where $\vartheta$ and $\varphi$ are polar and azimuthal angle of \vk~
respectively. Recent thermal conductivity experiments\cite{Izawa02c}
suggest that crystallographic [100] and [010] are the nodal
directions, i.e. the order parameter of Eq.~\ref{GAP} is rotated by
$\frac{\pi}{4}$ in
the a-b plane. This gap function accounts for the $\sqrt{H}$
dependence of the specific heat and the H- linear term in the thermal
conductivity observed recently\cite{Nohara99,Izawa01a,Boaknin01}. The
aim of this paper is to
generalize an earlier result\cite{Maki02} for $\kappa_{zz}$, $\kappa_{xx}$
and  $\kappa_{xy}$ for general magnetic field (\vH) orientation given by
the polar angle $\theta$ with respect to the c- axis and the azimuthal angle
$\phi$. First
in the absence of \vH~ the specific heat and the
electronic thermal conductivity for $\Gamma\ll T\ll \Delta$ are given by

\begin{eqnarray}
\frac{C_s}{\gamma_NT}&=&\frac{27}{4\pi}\zeta(3)(\frac{T}{\Delta})+..
\nonumber\\
\frac{\kappa_{xx}}{T}&=&\frac{\pi^2}{8}\frac{n}{m\Delta};\;\;\;\;
\frac{\kappa_{zz}}{T}=\frac{\pi^2}{8}\frac{nC_0}{m\Delta}
\end{eqnarray}

where
C$_0$=$(\frac{2\Gamma}{\Delta})^\frac{1}{2}
[\ln(2\sqrt{\frac{\Delta}{\Gamma}})]^{-\frac{1}{2}}$. Note that
$\kappa_{xx}$ obeys the universal behaviour while $\kappa_{zz}$ does not.
This is because the heat current operator j$_z^h$ vanishes on the
four second order nodal points $(\vartheta,\varphi)$ =
($\frac{\pi}{2},\pm\frac{\pi}{4}$) and
($\frac{\pi}{2},\pm\frac{3\pi}{4}$) for \De given in
Eq.(\ref{GAP}). Also this leads to a
H$^\frac{3}{2}\ln\Bigl(\frac{\Delta}{\tilde{v}\sqrt{eH}}\Bigr)$
dependence of $\kappa_{zz}$ as discussed below.\\

In the presence of a magnetic field with general orientation defined by
($\theta,\phi$) the specific heat and thermal conductivities in the
vortex phase are given by\cite{Maki02,Won01a}

\begin{eqnarray}
\label{COND}
\frac{C_s}{\gamma_NT}&=&
\frac{\tilde{v}\sqrt{eH}}{\sqrt{2}\Delta}I_+(\theta,\phi)\nonumber\\
\frac{\kappa_{zz}}{\kappa_n}&=&
\frac{1}{32\sqrt{2}}\ln\Bigl[\frac{2\Delta}{\tilde{v}\sqrt{eH}}
\sqrt{\frac{2}{1+\cos^2\theta}}\Bigr]
\frac{\tilde{v}^3(eH)^\frac{3}{2}}{\Delta^3}I_{zz}(\theta,\phi)\nonumber\\
\frac{\kappa_{xx}}{\kappa_n}&=&
\frac{3}{32}\frac{\tilde{v}^2(eH)}{\Delta^2}I_+^2(\theta,\phi)\\
\frac{\kappa_{xy}}{\kappa_n}&=&
-\frac{3}{32}\frac{\tilde{v}^2(eH)}{\Delta^2}I_-(\theta,\phi)I_+(\theta,\phi)
\nonumber
\end{eqnarray}

where we have the identity

\begin{eqnarray}
\label{S+GWAVE}
I_{zz}(\theta,\phi)&=&(1+\cos^2\theta)I_+(\theta,\phi)\nonumber\\
I_\pm(\theta,\phi)&=&\frac{1}{2}
\{[1+\cos^2\theta+\sin^2\theta\sin(2\phi)]^\frac{1}{2}\\
&&\pm\{[1+\cos^2\theta-\sin^2\theta\sin(2\phi)]^\frac{1}{2}\}
\nonumber
\end{eqnarray}

\begin{figure}
\centerline{\psfig{figure=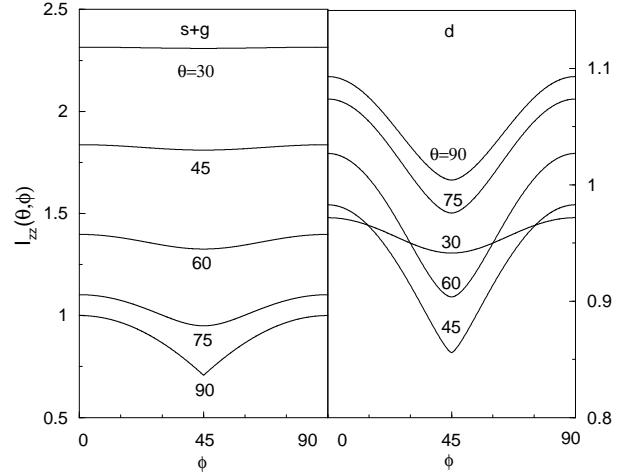,width=8cm}}
\vspace{0.5cm}
\caption{Angular dependence of I$_{zz}(\theta,\phi)$ which determines
$\kappa_{zz}(\theta,\phi)$ in the superclean limit for s+g wave and d-
wave case (up to log- terms in Eq. 3). Note the different
scale in the two cases.}
\label{FIGzz}
\end{figure}

\begin{figure}
\centerline{\psfig{figure=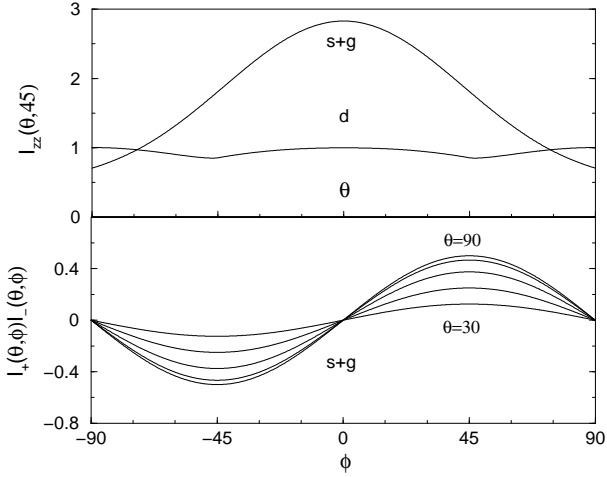,width=8cm}}
\vspace{0.5cm}
\caption{upper panel: polar angle variation of I$_{zz}(\theta$,45) for
s+g- and d- wave case. lower panel: 
Angular dependence of the product I$_-(\theta,\phi)\cdot$
I$_+(\theta,\phi)$ for s+g case which determines the thermal Hall coefficient
$\kappa_{xy}(\theta,\phi)$. It vanishes for field halfway between
the nodal directions due to current compensation.}
\label{FIGxy}
\end{figure}

Here we have assumed the superclean limit defined by
$\sqrt{\Delta\Gamma}\ll\tilde{v}\sqrt{eH}$ with
$\tilde{v}=\sqrt{v_av_c}$ where $v_{a,c}$ denote the anisotropic Fermi
velocities. The angular dependences of $\kappa_{zz}$ and $\kappa_{xy}$
according to Eqs.~\ref{COND},\ref{S+GWAVE} for the s+g wave case
are shown in the left panel of Fig.\ref{FIGzz} and in
Fig.\ref{FIGxy}. For comparison, we also present
the corresponding angular dependence of $\kappa_{zz}$ for the
d$_{x^2-y^2}$ state with \De=$\Delta\cos(2\phi)$ as in high T$_c$
cuprates\cite{Won01a}, CeCoIn$_5$\cite{Izawa01b} and
$\kappa$-(ET)$_2$Cu(NCS)$_2$ \cite{Izawa02a,Won01b}. Of course for d- 
wave superconductors the universal zero- field behaviour is valid both for
$\kappa_{xx}$ and for $\kappa_{zz}$, and both exhibit a similar
angular dependence in the vortex phase\cite{Won01a}. In this case the
dependence on field angles $\theta,\phi$ is given by

\begin{eqnarray}
\label{DWAVE}
I_\pm(\theta,\phi)&=&
\frac{1}{2\pi}\int_{-\frac{\pi}{2}}^{\frac{\pi}{2}}d\psi J_\pm(\psi)\nonumber\\
\tilde{I}_+(\theta,\phi)&=&
\frac{1}{2\pi}\int_{-\frac{\pi}{2}}^{\frac{\pi}{2}}d\psi 
(1-\cos(2\psi))J_+(\psi)\nonumber\\
J_\pm(\psi)&=&
\Bigl[1+\frac{1}{2}\sin^2\theta(\sin(2\phi)-\cos(2\psi))\\
&&+\frac{1}{\sqrt{2}}\sin(2\theta)\sin\psi\sqrt{1-\sin(2\phi)}\Bigr]
^\frac{1}{2}
\nonumber\\
&&\pm\Bigl[1-\frac{1}{2}\sin^2\theta(\sin(2\phi)+\cos(2\psi))\nonumber\\
&&+\frac{1}{\sqrt{2}}\sin(2\theta)\sin\psi\sqrt{1+\sin(2\phi)}\Bigr]
^\frac{1}{2}\nonumber
\end{eqnarray}

Then $\kappa_{xx}$ and $\kappa_{xy}$ are obtained from
I$_\pm(\theta,\phi$) as in Eq.~\ref{COND} but now for d- wave:

\begin{eqnarray}
\frac{\kappa_{zz}}{\kappa_n}&=&
\frac{1}{\pi}\frac{\tilde{v}^2(eH)}{\Delta^2}I_{zz}(\theta,\phi)\nonumber\\
I_{zz}(\theta,\phi)&=& I_+(\theta,\phi)\tilde{I}_+(\theta,\phi)
\end{eqnarray}

The $\phi$- dependence of $\kappa_{zz}$ for various $\theta$ is shown
in comparison to the s+g wave case in Fig.~\ref{FIGzz}. 
As is readily seen from Fig.~\ref{FIGzz} in the s+g- wave
case a pronounced cusp like feature develops for $\theta$=90$^\circ$
and $\phi=\pm$ 45$^\circ$ due to the (second order) point node, while in the
d$_{x^2-y^2}$ wave case with an extended line node along c no cusps
appear and also the absolute value of
angular variation is much smaller. This is clearly visible from the
upper panel of Fig.~\ref{FIGxy} which also shows monotonic $\theta$-
dependence for s+g- wave and nonmonotonic behaviour for d-wave. The
latter has a minimum at $\theta_m\simeq$ 47$^\circ$ which is due to a
maximum Doppler shift for $\theta$=45$^\circ$ resulting in a dominating term
I$_{zz}(\theta,\phi)\simeq 1-(5/64)\sin^2(2\theta)$+.. . Note that
I$_{zz}(\theta,\phi)$ in Fig.~\ref{FIGzz} exhibits a rather sharp
minimum as function of $\phi$ at $\theta_m$ wheras for $\theta$=
90$^\circ$ the minimum is flat. Experimentally however the $\kappa_{zz}$
thermal conductivity shows very strong cusps at $\theta$=90$^\circ$
(and $\phi$=0 due to rotated order parameter) in
\Y\cite{Izawa02c}. This is a strong point for
the s+g wave case being the appropriate one for \Y and \Lu. Therefore
the thermal conductivity in the superclean limit can discriminate s+g-
wave against d- wave superconductivity.

The angular dependence of the thermal Hall coefficient $\kappa_{xy}$
in the s+g wave case  is shown in the lower panel of
Fig.~\ref{FIGxy}. It exhibits a sign change as function of $\phi$ and
varies smoothly with $\theta$. In the d- wave case $\kappa_{xy}$ looks
rather similar.\\
As we have already discussed elsewhere\cite{Thalmeier02} the thermal
conductivity provides a unique window to look at the nodal structure of 
\De in unconventional superconductors.\\

\noindent
{\em Acknowledgement}\\[0.5cm] 
We would like to thank Koichi Izawa and Yuji Matsuda for useful
discussions on superconducting borocarbides.

\end{document}